\documentclass[prb,reprint]{revtex4-1} 
\usepackage{amsmath} 
\usepackage{graphicx} 
\usepackage{subfigure}

\begin{document}

\title{Scanning capacitance microscopy using a relaxation oscillator}
\author{M. Pahlmeyer}
\author{A. Hankins}
\author{S. Tuppan}
\author{W. J. Kim}
\email{kimw@seattleu.edu}
\affiliation{901 12th Ave Department of Physics, Seattle University, Seattle, WA 98122, USA}
\date{\today}

\begin{abstract}
We have performed scanning capacitance microscopy using a relaxation oscillator. Precision calibrations indicate a sensitivity on the order of 0.05 pF. Surface topography of metallic structures, such as machined grooves and coins, can be readily obtained either in the constant-height (non-contact) or tapping (contact) mode. Spatial resolution of sub-50 $\mu$m has been achieved. 
\end{abstract}

          
\maketitle

\section{Introduction}
Scanning capacitance microscopy (SCM) is a useful imaging technique that allows one to acquire topographical features of micro- and nano-sized samples. One of the early SCM instruments, based upon a commercial product (RCA CED video Disc),\cite{mat} is made of a capacitive sensor driven by an ultra-high frequency oscillator (500 MHz or higher), a sharp tip, and a particular sample to be imaged, all of which are in a feedback loop to maintain a maximum resonant output. Primary applications include: surface characterizations of nano structures\cite{will,step} and the profiling of both conductors and insulators, particularly to acquire semiconductor ($p$- and $n$-type) dopant density profiles,\cite{will2,gian} with sub-100 nm resolution. More recently, researchers have developed an integrated capacitance bridge for enabling quantum capacitance measurements at room temperature.\cite{haze} 

SCM is essentially a near-field capacitive sensor providing either the direct tip-sample capacitance ($C$), or its gradient with respect to the change of a tip-sample separation ($dC/dx$), the latter being a more common choice for imaging due to its high sensitivity achieved by a lock-in technique. Therefore, it is easy to recognize SCM as a variant of atomic force microscopy (AFM), since the tip-sample {\em electric capacitance} is directly related to the tip-sample {\em electric force}, which is proportional to the gradient of the capacitance itself. For this reason, SCM is frequently employed as another operating mode of atomic force microscopes, providing a direct measure of classical electrostatics between a tip and the surface of a sample.

Previously, we have reported the usefulness of a relaxation oscillator for measuring the capacitance between two metal plates in an attempt to characterize the absolute separation between them.\cite{hankins} Here, we extend our capacitance measurements to a topological mapping of the surface of a sample in both contact and non-contact modes. Our simple, low-cost SCM provides a valuable experimental platform in the undergraduate laboratory where students gain critical exposure to nano-scale imaging techniques. We present precision calibrations of the relaxation oscillator, and successful 2-D surface scans of machined grooves and an American coin. 

\section{Relaxation oscillator}
The relaxation oscillator, shown in the top of Fig. 1, consists of a comparator, an external capacitance to be measured, and three resistors. In our previous application,\cite{hankins} there was an internal capacitor $C_{\rm{int}}=47$ pF connected to $C_{\rm{ext}}$ in parallel; however, in the present oscillator, only the capacitance of an external source---plus possible parasitic capacitance---contributes to the characteristic oscillation period set by the $RC$ constant, where $R$ is the 100-k$\Omega$ resistor above the negative input of the op-amp. The other two 100-k$\Omega$ resistors, just below the positive input of the op-amp, act as a voltage divider, with $V_{\rm{+}}$ being just half of the maximum output of the oscillator $V_{\rm{out}}$. The total `capacitance' of the circuit, combining both $C_{\rm{ext}}$ and parasitic capacitance, repeats charging-discharging cycles at $V_{\rm{-}}$, which result in square waves at the oscillator output, as described in the bottom of Fig. 1. 
\begin{figure}[htbp]
\centering
\includegraphics[width=0.8\columnwidth,clip]{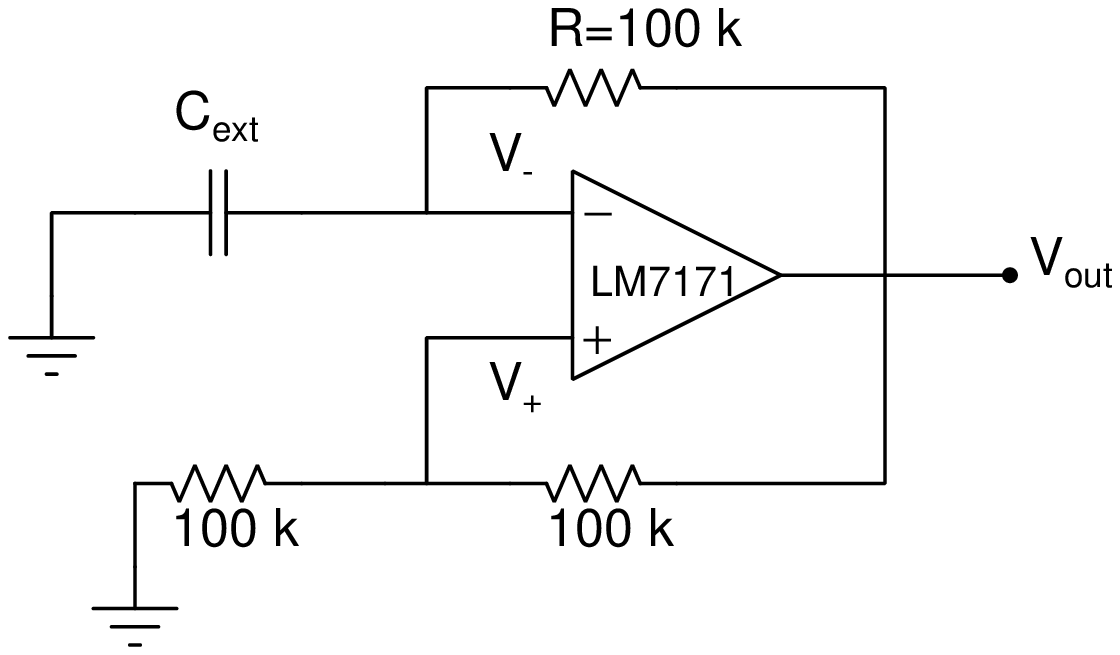}
\includegraphics[width=0.8\columnwidth,clip]{fig1b.eps}
\vspace{0.5cm}
\caption{Circuit diagram of a relaxation oscillator (top) and a plot of the oscillator and capacitor outputs, $V_{\rm{out}}$ and $V_{\rm{-}}$, respectively, captured by an oscilloscope (bottom). The total capacitance can be expressed as $C_{\rm{tot}}=C_{\rm{ext}}+C_{\rm{para}}$, where $C_{\rm{para}}$ represents possible parasitic capacitance present in the external wiring. Together with the 100 k$\Omega$-resistor, the $RC$ charging-discharging cycle is established. Note that the maximum of the capacitor output is half of $V_{\rm{out}}$ according to our 1:1 voltage divider.}
\label{fig1}
\end{figure}

To understand the circuit qualitatively, let us assume that $V_{\rm{out}}$ is initially held at a constant maximum voltage $V_0$. At this point, $V_{+}$ is exactly ${\frac{1}{2}V_{0}}$. The external capacitor then starts charging up and the voltage increases at $V_{-}$. Note that, as long as $V_{-}<V_{+}$, the output remains constant at $V_{0}$, and that the voltage across the capacitor keeps on increasing until $V_{-}>V{+}$, at which point the oscillator output swings to $-V_{0}$. Once the oscillator output flips, the capacitor starts discharging. As long as the output remains at $-V_{0}$, the capacitor continues to discharge until $V_{+}>V_{-}$. This cycle repeats itself. 

Quantitatively, we can derive the relationship between the period of the oscillation and the $RC$-time constant.\cite{ham} Applying Kirchhoff's rule and Ohm's law to the $RC$ components of the op-amp, we get: $V_{\rm{out}}=V_{-}+V_{\rm{R}}$, where $V_{\rm{R}}$ is the voltage drop across the resistor. This leads to
\begin{equation}
V_{\rm{out}}-V_{-}=RC\frac{dV_{-}}{dt},
\end{equation}  
where $C$ is the total capacitance of the system and $R$ is the 100 k$\Omega$-resistor. The period of the oscillation $T$ can be calculated by considering one half-period during which the capacitor output at $V_{-}$ swings from $-\frac{1}{2}V_0$ to $+\frac{1}{2}V_0$. Solving Eq. 1 with initial conditions: (i) $V_{-}=-\frac{V_0}{2}$; and (ii) $V_{\rm{out}}=V_0$ at $t$=0 yields
\begin{equation}
V_{-}=V_0(1-\frac{3}{2}e^{-t/RC}).
\end{equation}
The time elapse, when $V_{-}$ reaches $+\frac{1}{2}V_0$, is exactly a half-period (e.g. $t=T/2$), so the period of the oscillation is immediately obtained as
\begin{equation}
T=2\ln(3)RC.
\end{equation} 
A few remarks are in order with regard to Eq. 3. First, the period of the relaxation oscillator is directly proportional to the capacitance being measured. To give a rough estimate, for $C=10$ pF and $R=100$ k$\Omega$, a typical period to be measured is on the order of 10 $\mu$s (see Section III). Second, Eq. 3 contains the DC offset (i.e. $T_0$ or baseline value of $T$) due to the parasitic capacitance. That is, $T=2\ln(3)R(C_{\rm{ext}}+C_{\rm{para}})$, where the total capacitance is the sum of the {\em variable} external capacitance and the {\em static} parasitic capacitance originating from the circuit wiring and other contacts. Finally, to obtain rough estimates of resolution related to differential capacitance, we take the inverse of the derivative of Eq. 3 with respect to $C_{\rm{ext}}$
\begin{equation}
\frac{\partial{C_{\rm{ext}}}}{\partial{T}}=\frac{1}{2\ln(3)R},
\end{equation}
which gives approximately 5 pF/$\mu$s resolution. For a $100$-MHz (or 0.01-$\mu$s) counter, the lower bound of capacitance sensitivity is 0.01 pF, which is compatible or even better than some commercially available RCL meters. In reality, the actual sensitivity is limited by drifts in measurements, such as those caused by temperature and acoustic vibrations, as noticed in our own measurements (see Section IV). 

In selecting an op-amp, slew rate is an important consideration as the oscillator output rapidly swings between two extrema (e.g. positive to negative). We have chosen LM7171 as a comparator due to its fast slew rate (4100 V/$\mu$s) and low-noise performance (14 nV/$\sqrt{\rm{Hz}}$). Our previous choice OP27, despite its lower noise-performance (3 nV/$\sqrt{\rm{Hz}}$), has a slew rate of only about 3 V/$\mu$s, which overrides the lower bound of capacitance set by the counter. Other options, such as  AD790 and AD8561 that are specifically designed to operate as comparators, provide ultra-fast rise and fall time on the order of nano-seconds. Another approach for building a relaxation oscillator has been proposed by Liu {\textit {et al}}..\cite{liu} In their configuration, the oscillator operates on two op-amps, one as an integrator for active measurements of the $RC$ constant and the other as a comparator to measure the charging-discharging cycle. While the two approaches are equally viable, we have chosen the single-comparator configuration because of its stable outputs and its simplicity.  

\section{Experimental setup and calibrations}
To perform scanning capacitance microscopy, a stage is constructed with three mechanical actuators (two Z825BV and one Z812BV from Thorlabs), which move in the three Cartesian axes, as shown in Fig. 2. Each actuator has 50 nm resolution and travel ranges of 25 mm and 12 mm in lateral and vertical translations, respectively. It is important to ensure good insulation and rigid placement of various parts in the vicinity of tip and sample, because the relaxation oscillator is found to be extremely delicate. For example, it can be disturbed by a tiny capacitance as a result of walking near the circuit. We have used a glass tube (Vitro tube) to secure the probe tip, and the various wires connecting to the relaxation oscillator are kept as short as possible to further minimize interference.  
 \begin{figure}[htbp]
\centering
\includegraphics[width=1.0\columnwidth,clip]{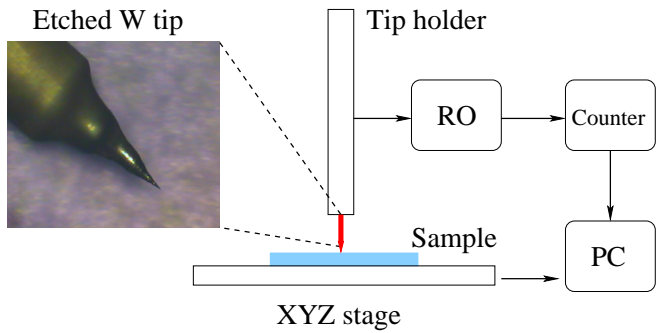}
\caption{Schematic of the experimental setup: The XYZ stage is controlled by the PC to move in relation to the probe tip which is held at a constant location in space. The sample and the probe tip form $C_{\rm{ext}}$ and consequently determine the period of the relaxation oscillator (RO). The period is measured by the counter (Keithley 2015 THD Multimeter) and collected by the PC via a MATLAB data acquisition program. The etched tungsten has a tip size on the order of 2 $\mu$m, which can be readily estimated from the initial diameter of the non-etched wire of 500 $\mu$m. See inset.}
\label{fig2}
\end{figure}

The probe tip is electrochemically etched on a tungsten wire in 2 M sodium hydroxide inside a teflon cell.\cite{kim,mcd} A positive electrode (+2 V) is connected to the top of the tungsten wire with a negative electrode (ground) in the solution. The electrical energy coupled with the basic solution induces effective etching during which the wire surrounded by the meniscus is etched slower than the wire deeper in the solution. Because of this, the tungsten wire becomes thinner towards the tip. The process continues until the weight of the base of the wire exceeds the tensile strength of the etched portion of the wire. When the base falls off, it leaves a sharp needle-like tip, as shown in Fig. 2. Alternatively, a paperclip can be used as a probe tip by cutting end diagonally, which also produces a relatively sharp tip point. 

To calibrate the relaxation oscillator, we used a number of precision capacitors (Vishay Sprague 5\% tolerance), across a range of capacitance values from 1 pF to 100 pF. This direct calibration, shown in the top of Fig. 3, gives a calibration factor of $\alpha=0.206\pm0.032$ $\mu$s/pF (or 4.87$\pm0.75$ pF/$\mu$s) with $T_0=5.2\pm1.0$ $\mu$s, which enables a conversion of period in seconds (s) to the actual unit of capacitance (F). From the standard deviation of the mean (SDOM) of $N$=50 measurements, our oscillator reaches a resolution of 0.001 pF in about 3 seconds. We have also confirmed the capacitance of the precision capacitors using an RCL meter (Fluke PM6303A).
\begin{figure}[htbp]
\centering
\includegraphics[width=1.0\columnwidth,clip]{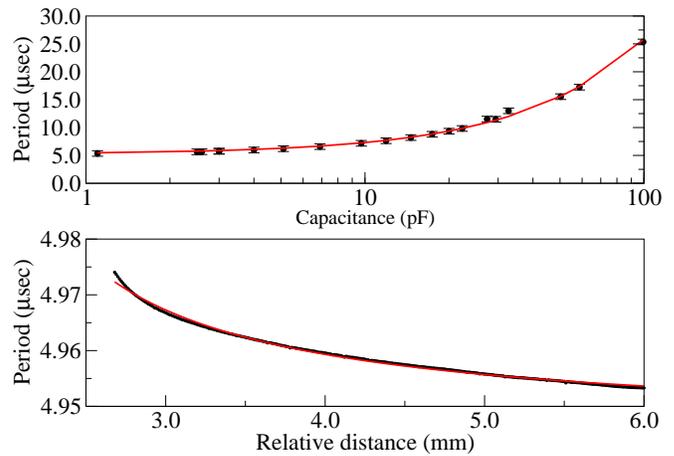}
\caption{Direct calibration using precision capacitors of known capacitance (top) and the parallel-plate calibration by mapping the periods at different separation distances between the probe tip and the sample surface (bottom). By fitting the data from the direct calibration to a linear function (i.e. first-order polynomial), we obtain a period-to-capacitance calibration factor: $\alpha$=0.206 $\mu$s/pF. The parallel-plate calibration is somewhat imprecise, because the tip-sample geometry is not precisely known {\em a priori}. The deviation from the $1/d$ fit is evident as the tip-probe separation is decreased.}
\label{fig3}
\end{figure}

An alternative way to calibrate the relaxation oscillator is to employ a parallel-plate system in which capacitance is expected to obey $C=\epsilon_0A/d$ (with $A$ being the effective area of the system) and then to fit the resulting data with a power law form: $T=T_0+\beta/(d-d_0)^{\gamma}$, where $T_0$ is the offset period,  $d_0$ is a point of contact, and $\gamma=1$. The pre-factor, $\beta$, is directly related to physical parameters, such as $\epsilon_0$ and $A$. The bottom of Fig. 3 is generated by first bringing the tip almost in contact with the sample, in this case approximately 100 $\mu$m, and then stepping further away in increments of 1 $\mu$m until it reaches a maximum distance determined by the range of the linear actuator. The graph, though strongly displaying the $1/d$ relationship at larger distances (i.e. away from the other plate), suffers a severe deviation from the expected power law at short distances, possibly, due to finite parallelism. For this reason, the $1/d$ calibration is not as reliable as the direct calibration using capacitors of known capacitance. From the fit, we obtain: $T_0=4.946\pm0.001$ $\mu$s, $d_0=1.332\pm0.001$ mm, and $\beta=0.035\pm0.001$ $\mu$s$\cdot$mm (or $0.039\pm0.001$ pF$\cdot$mm using the calibration factor $\alpha$). Since $\beta\equiv\epsilon_0A$, we estimate the effective diameter of the probe tip to be roughly $D_{\rm{eff}}=(2.2\pm0.1$) mm. This is about a factor of three larger than the actual diameter of the probe tip measured to be $D_{\rm{act}}=(0.8\pm0.1)$ mm. In fact, this discrepancy is not surprising, because the precise geometry of the probe tip cannot be known {\em a priori} and, as a consequence, validity of the parallel-plate model does not hold (i.e. $D_{\rm{act}}\neq D_{\rm{eff}}$). The calibration involving a parallel-plate system is an interesting subject in its own right and enables critical assessments of a finite degree of parallelism manifested by a possible deviation from the expected power law.\cite{hankins,mcd} 

\section{Temperature drifts}
We have found a striking correlation between capacitance and temperature. The tip-sample capacitance at a fixed position, as measured by the period of the relaxation oscillator, is shown to be inversely proportional to the temperature of the op-amp, as measured by the resistance of a 10-k$\Omega$ thermistor (TH10k from Thorlabs). This behavior is clearly seen in Fig. 4. The dotted line corresponds to the resistance of a thermistor mounted inside the relaxation oscillator circuit box during an 11-day period. The actual relationship between the resistance of the thermistor and temperature is rather complicated.\cite{thorlab} But, to give a rough estimate, we have $\Delta T_{\rm{C}}/\Delta R\sim$2 C$^{\circ}$/k$\Omega$ at room temperature ($\sim20 ^\circ$C), translating into 2 C$^{\circ}$ change over the course of the 11-day period. This variation in temperature then corresponds to $\Delta C=0.3 $ pF change in capacitance.
\begin{figure}[htbp]
\centering
\includegraphics[width=1.0\columnwidth,clip]{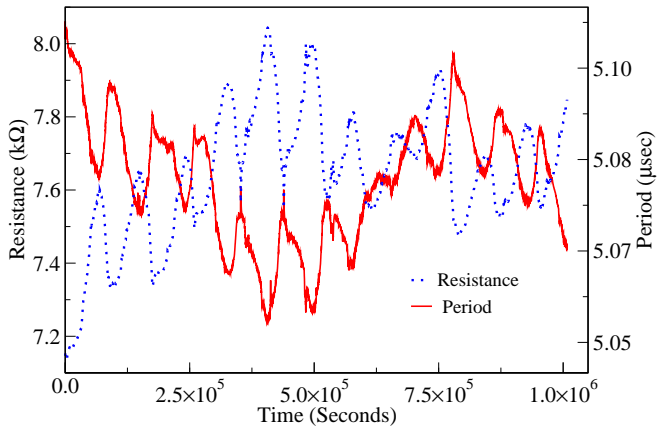}
\caption{Plot of thermistor resistance (i.e. temperature) and period (i.e. capacitance) data over the course of 11 days displaying a strong correlation between them: As a result, a small drift of room temperature strongly affects the capacitance measurements. For temperature measurements, we have used a thermistor whose resistance is inversely proportional to temperature. Hence, the actual temperature and the capacitance are proportional to each other (e.g. positively correlated).}
\label{fig4}
\end{figure}

To suppress the temperature-driven fluctuations, it is possible to implement a temperature control system in which one side of the op-amp is in contact with a thermoelectric coupler (TEC), while the other side is in contact with a heat sink. A proportional-integral-derivative (PID) controller then actively stabilizes the system to a set temperature. Building a PID controller circuit is relatively simple and provides great stability ($\sim10$ mK).\cite{caltech} 

In the absence of a temperature-stabilizing mechanism, it becomes necessary to quantify the degree of correlation to account for any temperature-driven capacitance drift. 
\begin{figure}[htbp]
\centering
\includegraphics[width=1.0\columnwidth,clip]{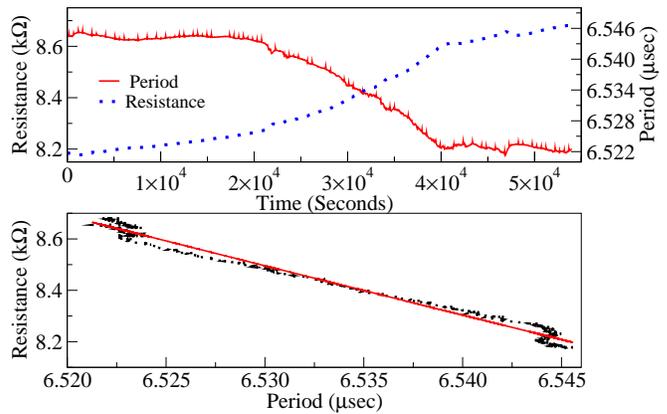}
\caption{A closer look of the temperature-capacitance correlation over a 15-hour period: The top graph plots both resistance and period over time to show the inverse relationship between them, while the lower graph plots them against each other to extract the linear correlation coefficient, which is calculated to be $\vert r\vert=0.99$ with a slope of 18.5 k$\Omega$/$\mu$s. The variation in period during 15 hours corresponds to an overall change in capacitance of 0.05 pF. }
\label{fig5}
\end{figure}
Because our typical surface scan (in the non-contact mode) takes on the order of several hours, we focus on the capacitance-temperature fluctuation over a 15-hour period (top of Fig. 5) and obtain the linear correlation coefficient (bottom of Fig. 5) using the definition\cite{taylor}: $r\equiv\frac{\sigma_{xy}}{\sigma_x\sigma_y}$, where $\sigma_{xy}=\sum(x_i-\bar{x})(y_i-\bar{y})/N$ is the covariance of $x$ (period) and $y$ (resistance) and $\sigma_x=\sqrt{\sum(x_i-\bar{x})^2}$ and $\sigma_y=\sqrt{\sum(y_i-\bar{y})^2}$ represent the standard deviation of $x$ and $y$, respectively, with each sum running over $N$ measurements. The obtained linear correlation coefficient is 0.99 with a best slope of 18.5 k$\Omega$/$\mu$s, or in terms of the actual unit of capacitance and temperature: 0.2 pF/C$^{\circ}$. This slope can be then directly used to correct any temperature-driven fluctuations.

\section{Scanning result I: Non-contact mode}
Presented in Fig. 6 is the result of scanning over a sample of brass machined `S' and `U' (for Seattle University). The two letters are convex-shaped with depth of 2 mm. For this particular scan, a paperclip is used as a probe tip and moves across the whole sample at a constant height in two dimensions (5 mm by 4 mm). Hence, the variation in capacitance is directly caused by the change in depth in the sample itself and represents a topographical map of the sample surface. Data are collected approximately every 100 $\mu$m for a total of 7200 points. 
\begin{figure}[htbp]
\centering
\includegraphics[width=1.0\columnwidth,clip]{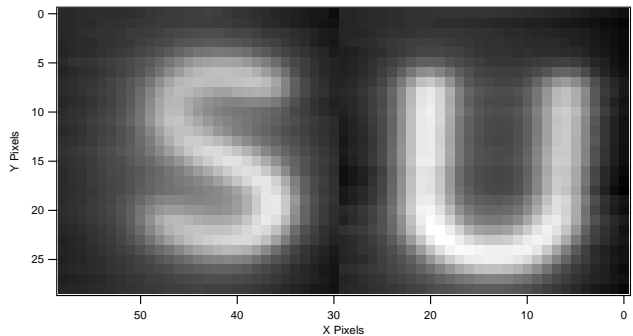}
\caption{Surface plot of a two-dimensional scan in constant-height mode: The probe tip is held at a fixed height and the stage containing the sample is moved in relation to it. Capacitance (period) measurements are conducted at regularly spaced intervals. The capacitance data directly correspond to height variations and subsequently produces a topographical map of the sample being scanned. The distance between pixels (i.e. lateral resolution) is 100 $\mu$m. A clear contrast of depth is obtained with the letters `SU' distinctly resolved.}
\label{fig6}
\end{figure}

The main advantage of this method is that the tip and the sample never come in direct contact. Reliable scans are obtained as long as the tip-sample distance is maintained sufficiently small ($<$10 $\mu$m). Although the method is completely ``touch-free'' and thus non-destructive, several disadvantages also exist: First, as we have previously discussed, the effect of the temperature drift over time causes the capacitance to fluctuate. The capacitance fluctuation is quite inevitable in the non-contact method, and either active temperature stabilization or off-line treatment of data using the linear correlation coefficient becomes necessary. Second, the system is extremely susceptible to acoustic disturbance. For example, the period measurement can be affected by something as minor as a person walking within a few feet of the experimental setup. It might be useful to place the setup inside an enclosure on a vibration-control table, but our scanning setup sat on an optical table without any enclosure. Finally, it is significantly less capable of scanning sharp cliffs or valleys in a sample, because the probe tip is not only interacting with the section of the sample immediately below it, but also with portions of the sample in the immediate vicinity of the relevant point. Depending on the topography of the sample, the surrounding (secondary) portion of the sample may be much closer to the tip than the relevant (primary) portion of the sample to be imaged, leading to an inflated period measurement.

\section{Scanning result II: Contact mode}
The contact method, a product of which is shown in Fig. 7 as a two-dimensional scan over the face of Abraham Lincoln on a US penny, actually brings the probe tip into direct contact with the sample. In this contact method, we (i) record the height at which a sharp fall in period measurements occurs; and (ii) retract the tip a set distance away from the initial location, usually about 1 $\mu$m, and (iii) approach a new point to find the height at which the sharp fall occurs at that location. The procedure is repeated until a full, topological map of height variations is acquired over the entire surface of a sample. 
\begin{figure}[htbp]
\centering
\includegraphics[width=1.0\columnwidth,clip]{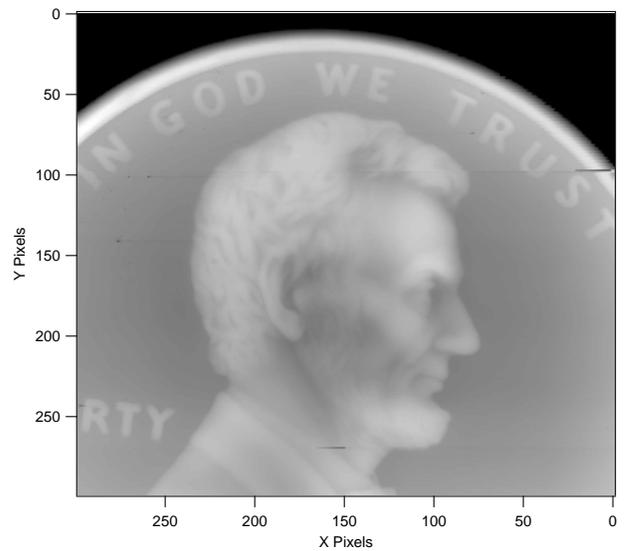}
\caption{Surface plot of a two-dimensional scan of a US penny in contact mode. At a given lateral position, the probe tip comes into physical contact with a point on the sample surface. The heights at which the capacitance (period) value rapidly falls are recorded and subsequently generate a topographical map of the sample. This method is somewhat slower, but yields a much better resolution, because it is insensitive to externally-driven fluctuations. The distance between each pixel is 45 $\mu$m for this run.}
\label{fig7}
\end{figure}

The main advantage to this method is that it is significantly more precise and reliable than the non-contact method. While the non-contact method is extremely vulnerable to external noises, the contact method is almost impervious to them, making it drift-free. In this method, the {\em contrast} resolution and {\em lateral} resolution are limited by the resolution of the linear actuation (50 nm) and the size of the employed tip (2 $\mu$m), respectively. However, one major disadvantage is that, owing to the probe tip making repeated contacts with the sample surface, the end tip will blunt over time and, consequently, the lateral resolution degrades as the scanning progresses. Additionally, delicate portions of the sample could be damaged by the light, yet frequent, touch of the probe tip. 

In principle, the capacitance measurement is not at all necessary to perform the contact microscopy. The only requirement is the capability to measure the contact transition when a sharp tip `touches' the surface of a sample. For this reason, a simple, battery-operated current-meter can be built to replace the present relaxation oscillator. There, the contrast resolution is still set by the minimum distance of the z-axis translation, and the lateral resolution still limited by the size of the probe tip.

\section{Conclusion}
We have reported results of scanning capacitance microscopy using a relaxation oscillator. Successful surface topography of both machined grooves and a coin has been obtained in the non-contact and contact modes with a spatial resolution of 100 $\mu$m and 45 $\mu$m, respectively. Our simple approach provides an excellent opportunity for students wishing to gain laboratory exposure to nano-scale imaging and microscopy techniques. To advance the present technique to a next level, the linear mechanical actuators can be replaced by a set of pico-meter piezoelectric translators (PZTs), and the tungsten probe tip can be attached with a template-assisted, electrochemically-synthesized nanowire whose diameter is as small as 10 nm.\cite{kim} Despite some apparent challenges, such as manipulating an individual nanowire for attachment and maintaining a sharp-contact point, the ultra-small tip, aided by high-precision PZTs, are within experimental reach and could drastically improve the presently achieved resolution by more than three orders of magnitude. 

\begin{acknowledgments}
This work is supported by the Clare Luce Boothe Research Program (MP), the M. J. Murdock Charitable Trust, Pat and Mary Welch (ST and AH), and the Research at Undergraduate Institutions through the National Science Foundation (WJK). We also thank Charlie Rackson for careful proofreading of our manuscript.   
\end{acknowledgments}

\end{document}